\documentclass[preprint]{revtex4}
\usepackage{amsmath}
\usepackage{graphicx}
\usepackage{amssymb}
\usepackage{amsthm, amscd}
\usepackage{amsfonts}
\usepackage{cancel}
\usepackage{times}
\usepackage{pstricks}
\usepackage{wasysym}
\usepackage{ulem}
\usepackage[dvips]{epsfig}
\usepackage{subfigure}
\usepackage{feynmf}

\usepackage{amssymb}

\newcommand{\dslash}{\not{\hbox{\kern-2pt $\partial$}}}
\newcommand{\td}{\tilde} 
\newcommand{\bq}{\begin{equation}} 
\newcommand{\eq}{\end{equation}}
\newcommand{\bqa}{\begin{eqnarray}} 
\newcommand{\eqa}{\end{eqnarray}}
\newcommand{\nn}{\nonumber \\}
\newcommand{\bw}{\begin{widetext}}
\newcommand{\ew}{\end{widetext}}

\newcommand{\JJ}{{\cal J}}
\begin{document}


\title{Holographic description of quantum field theory}

\author{Sung-Sik Lee}
\affiliation{Department of Physics $\&$ Astronomy, 
McMaster University,
Hamilton, Ontario L8S 4M1, 
Canada}

\date{\today}

\begin{abstract}

We propose that
general $D$-dimensional quantum field theories
are dual to $(D+1)$-dimensional local quantum theories 
which in general include objects with spin two or higher.
Using a general prescription,
we construct 
a $(D+1)$-dimensional theory 
which is holographically dual to
the $D$-dimensional $O(N)$ vector model.
From the holographic theory, 
the phase transition and critical properties of the model 
in dimensions $D>2$ are described.


\end{abstract}

\maketitle

\section{Introduction}

Quantum field theory is a universal language
that describes long wavelength fluctuations
in quantum systems made of many degrees of freedom.
Although strongly coupled quantum field theories
commonly arise in nature,
it is notoriously difficult to find 
a systematic way of understanding 
strongly coupled quantum field theories.

The anti-de Sitter space/conformal field theory correspondence\cite{MALDACENA,GUBSER,WITTEN}
opened the door to understand a class of strongly coupled
quantum field theories.
According to the duality,
certain strongly coupled quantum field theories 
in $D$ dimensions
can be mapped into
weakly coupled gravitational theories 
in $(D+1)$ dimensions
in large $N$ limits.
Although the original correspondence has been conjectured
based on the superstring theory,
it is possible that the underlying principle 
is more general
and a wider class of 
quantum field theories
can be understood 
through holographic descriptions\cite{KLEBANOV,DAS,Gopakumar:2004qb,POLCHINSKI09},
 which may have different UV completion than the string theory.

In this paper, we provide a prescription
to construct holographic theories 
for general quantum field theories.
As a demonstration of the method,
we explicitly construct a dual theory
for the $D$-dimensional $O(N)$ vector model, 
and reproduce the 
phase transition and critical behaviors 
of the model using the holographic description.

The paper is organized in the following way.
In Sec. II, we will convey the main idea behind the 
holographic description
by constructing a dual theory
for a toy model.
In Sec. III A, 
using the general idea presented in Sec. II,
we will explicitly construct a holographic theory dual to the 
$D$-dimensional $O(N)$ vector model.
In Sec. III B, 
we will consider a large $N$ limit
where the theory becomes classical
for $O(N)$ singlet fields in the bulk.
In Sec. III C, 
the phase transition and critical properties
of the $O(N)$ model will be discussed
using the holographic theory.

\section{Toy-model : $0$-dimensional scalar theory}

In this section, 
we will construct a holographic theory
for one of the simplest models : 
$0$-dimensional scalar theory.
In zero dimension, the partition function
is given by an ordinary integration, 
\bqa
Z[\JJ] & = & \int d \Phi ~e^{-S [\Phi]}.
\label{ZJ}
\eqa
We consider an action $S[\Phi] = S_M[\Phi] + S_\JJ[\Phi]$ with
\bqa
S_{M}[\Phi] &=& M^2 \Phi^2, \nn
S_\JJ[\Phi] &=& \sum_{n=1}^\infty \JJ_n \Phi^n.
\eqa
Here $S_M$ is the bare action with `mass' $M$.
$S_\JJ$ is a deformation with sources $\JJ_n$.
The values of $\JJ_n$'s are not necessarily small.
In the following,
we will consider deformations
upto quartic order : $\JJ_n = 0$ for $n>4$.
However, the following discussion can be 
straightforwardly generalized to
more general cases.

For a given set of sources $\JJ_n$,
quantum fluctuations are controlled
by the bare mass $M$.
One useful way of organizing quantum fluctuations
is to separate high energy modes and low energy modes,
and include high energy fluctuations 
through an effective action 
for the low energy modes.
Although there is only one scalar variable in this case, 
this can be done through 
the Polchinski's renormalization group scheme\cite{POLCHINSKI84}.
First, an auxiliary field $\td \Phi$ with mass $\mu$ is introduced,
\bqa
Z[\JJ] & = & \mu \int d \Phi d \tilde \Phi ~e^{-(S[\Phi] + \mu^2 \tilde \Phi^2)}.
\label{eq:ax}
\eqa
At this stage, $\tilde \Phi$ is a pure auxiliary field 
without any physical significance.
Then, we find a new basis $\phi$ and $\tilde \phi$ 
\bqa
\Phi &=& \phi + \tilde \phi, \nn
\tilde \Phi &=& A \phi + B \tilde \phi,
\eqa
in such a way that
the `low energy field' $\phi$ has a mass $M^{'}$
which is slightly larger than the original mass $M$.
As a result, quantum fluctuations for $\phi$ become
slightly smaller than the original field $\Phi$.
The missing quantum fluctuations are 
compensated by the `high energy field' $\tilde \phi$
with mass $m^{'}$.
If we choose the mass of the low energy field $\phi$ as 
\bqa
M^{'2} &=& M^2 e^{2 \alpha dz} 
\eqa
with $dz$ being an infinitesimally small parameter
and $\alpha$ being a positive constant,
we have to choose
\bq
A = -\frac{M M^{'}}{ m^{'} \mu}, ~~ B =  \frac{m^{'} M }{ M^{'} \mu},
\eq
where
\bqa
m^{'2} &=& M^2 \frac{ e^{2 \alpha dz} }{ e^{2 \alpha dz} - 1} = \frac{M^2}{ 2 \alpha dz}.
\eqa
Note that $m^{'2}$ is very large, 
proportional to $1/dz$.
This is because $\td \phi$ carries away
only infinitesimally small quantum fluctuations 
of the original field $\Phi$.
Moreover, $m^{'}$ is independent of 
the arbitrary mass $\mu$
because $\td \phi$ is physical.

In terms of the new variables, 
the partition function is written as
\bqa
Z[\JJ] & = &  \left( \frac{M m^{'}}{ M^{'}}  + \frac{M M^{'}}{ m^{'} }   \right) 
\int d \phi d \tilde \phi ~e^{-(S_\JJ[\phi + \tilde \phi] + M^{'2} \phi^2 + m^{'2} \tilde \phi^2)}.
\eqa
If we rescale the fields,
\bqa
\phi  \rightarrow  e^{- \alpha dz } \phi, ~~~~ 
\td \phi  \rightarrow  e^{- \alpha dz } \td \phi,
\eqa
the quadratic action for low energy field $\phi$
can be brought into the form 
which is the same as 
the original bare action, 
\bqa
Z[\JJ] & = &  m  
\int d \phi d \tilde \phi ~e^{-(S_j[\phi + \td \phi]  + M^2 \phi^2 + m^2 \tilde \phi^2)},
\eqa
where 
\bqa
S_j[\phi + \td \phi] &=& \sum_{n=1}^4 j_n ( \phi + \td \phi)^n, 
\eqa
with $j_n = \JJ_n e^{- n \alpha dz }$ 
and $m = m^{'} e^{-\alpha dz}$.
Note that $j_n$'s become smaller than the original sources $\JJ_n$,
which is a manifestation of reduced quantum fluctuations
for the low energy field $\phi$.
The new action can be expanded in power of the low energy field,
\bqa
S_j[\phi + \td \phi] &   = & S_j[\td \phi]  
 + ( j_1 + 2 j_2 \td \phi + 3 j_3 \td \phi^2 + 4 j_4 \td \phi^3 ) \phi \nn
&& + ( j_2 + 3 j_3 \td \phi + 6 j_4 \td \phi^2 ) \phi^2 
 + ( j_3  + 4 j_4 \td \phi ) \phi^3 
 + j_4 \phi^4. 
\label{expansion}
\eqa
In the standard renormalization group (RG) procedure\cite{POLCHINSKI84,POLONYI},
one integrates out the high energy field 
to obtain an effective action for the low energy field
with renormalized coupling constants.
Here we take an alternative view 
and interpret the high energy field $\td \phi$ 
as fluctuating sources for the low energy field.
This means that the sources for the low energy field 
can be regarded as dynamical fields 
instead of fixed coupling constants. 
To make this more explicit, 
we decouple the high energy field and the low energy field
by introducing Hubbard-Stratonovich fields $J_n$ and $P_n$,
\bqa
Z[\JJ] & = &   m  
\int d \phi d \tilde \phi  \Pi_{n=1}^4 (d J_n d P_n ) ~ 
e^{-(S_j^{'} + M^2 \phi^2 + m^2 \tilde \phi^2)},
\eqa
where
\bqa
S_j^{'} & = &  S_j[ \td \phi]  \nn
&& + i P_1 J_1 - i P_1 ( j_1 + 2 j_2 \td \phi + 3 j_3 \td \phi^2 + 4 j_4 \td \phi^3 ) + J_1 \phi \nn
&& + i P_2 J_2 - i P_2 ( j_2 + 3 j_3 \td \phi + 6 j_4 \td \phi^2 ) + J_2 \phi^2 \nn
&& + i P_3 J_3 - i P_3 ( j_3  + 4 j_4 \td \phi ) + J_3 \phi^3 \nn
&& + i P_4 J_4 - i P_4 j_4 + J_4 \phi^4.
\label{Si}
\eqa
Now we integrate out $\td \phi$ 
to obtain an effective action 
for the source fields.
The mass $m^2$ for the high energy field 
is proportional to $1/dz$ and 
only terms that are linear in $dz$
contribute to the effective action
(for the derivation, see the Appendix A),
\bqa
Z[\JJ] & = &  
\int d \phi \Pi_{n=1}^4 (d J_n d P_n ) ~ e^{-(S_{J}[\phi] + M^2 \phi^2 + S^{(1)}[J,P])},
\label{eq:dz}
\eqa
where
\bqa
S^{(1)}[J,P] 
&= &  \sum_{n=1}^4 i ( J_n - \JJ_n + n \alpha dz \JJ_n ) P_n  \nn
&& + \frac{\alpha dz}{2 M^2} ( i \td \JJ_1 + 2  P_1 \td \JJ_2 + 3  P_2 \td \JJ_3 + 4  P_3 \td \JJ_4 )^2
\label{S1}
\eqa
with
$\td \JJ_n = (\JJ_n + J_n)/2$.

After repeating the steps from Eqs. (\ref{eq:ax}) to (\ref{eq:dz}) 
$R$ times, 
one obtains a path integral for the partition function
\bqa
Z[\JJ] & = & \int \Pi_{k=1}^{R} \Pi_{n=1}^4 (D J_n^{(k+1)} D P_n^{(k)} ) 
e^{-S^{(R)}[J^{(k)},P^{(k)}]} Z[ J^{(R+1)} ],
\label{eq:recur}
\eqa
where
\bqa
S^{(R)}[J^{(k)},P^{(k)}] 
&= & \sum_{k=1}^{R} \Bigl[
\sum_{n=1}^4 i ( J_n^{(k+1)} - J_n^{(k)} + n \alpha dz J_n^{(k)} ) P_n^{(k)}  \nn
&& +\frac{\alpha dz}{2 M^2} ( i \td J_1^{(k)} + 2  P_1^{(k)} \td J_2^{(k)} + 3  P_2^{(k)} \td J_3^{(k)} + 4  P_3^{(k)} \td J_4^{(k)} )^2 
\Bigr] 
\label{Sgravity_d}
\eqa
with
$\td J_n^{(k)} = (J_n^{(k+1)} + J_n^{(k)})/2$
and $J_n^{(1)} = \JJ_n$.
The non-trivial solution for Eq. (\ref{eq:recur}) is given by
\bqa
Z[\JJ] & = &    
\int \Pi_{n=1}^4 (D J_n D P_n ) ~ e^{-S[J,P]},
\label{Zgravity}
\eqa
where
\bqa
S[J,P] 
&= & \int_0^\infty dz \Bigl[  i ( \partial_z J_n + n \alpha J_n ) P_n  \nn
&& + \frac{\alpha}{2M^2} ( i J_1 + 2  P_1 J_2 + 3  P_2 J_3 + 4  P_3 J_4 )^2  \Bigr]. 
\label{Sgravity}
\eqa
Here $DJ_n DP_n$ represent functional integrations 
over one dimensional fields $J_n(z), P_n(z)$ 
which are defined on the semi-infinite line $[0,\infty)$.
The boundary value of $J_n(z)$ is 
fixed by the coupling constants of 
the original theory, $J_n(0) = \JJ_n$.
$P_n(z)$ is the conjugate field of $J_n(z)$.
The physical meaning of $P_n$ becomes
clear from the equation of motion for the corresponding source field.
By taking derivative of Eq. (\ref{Si}) with respect to $J_n$,
one obtains
\bqa
< \phi^n > = - i < P_n >.
\eqa
Therefore, $P_n$ describes
physical fluctuations of the operator $\phi^n$, 
and an expectation value of $P_n$ 
along the imaginary axis gives 
an expectation value of the operator.

The theory given by Eqs. (\ref{Zgravity}) and (\ref{Sgravity}),
which is exactly dual to the original theory,
is an one-dimensional local quantum theory.
The emergent dimension $z$ corresponds to logarithmic energy scale\cite{VERLINDE}.
The parameter $\alpha$ determines the rate the energy scale is changed.
Despite the apparent similarity with 
the standard RG theory,
there is an important difference.
In the usual RG approach, 
quantum fluctuations of high energy modes
modify coupling constants for 
low energy modes
and generated new terms 
in the effective action.
In the present approach,
new terms are not generated
and the structure of the bare theory
is maintained at each level.
In particular, the highest order coupling ($J_4$ in this case)
obeys the strict constraint,
\bqa
 \partial_z J_4 + 4 \alpha J_4 = 0
\label{J4} 
\eqa
which is nothing but the classical scaling.
This is because there is no dynamics for the conjugate field $P_4$
for the highest order coupling.
Instead, other couplings acquire non-trivial dynamics
and some of them become propagating modes in the bulk.
Fluctuations of those dynamical {\it coupling fields} embody
quantum fluctuations in this approach.

In quantum field theory,
there is a redundancy
as to what energy scale one should use to define theory.
This makes the reparametrization of the RG flow 
to be a gauge symmetry.
Because of this redundancy,
the partition function in Eq. (\ref{Zgravity})
does not depend on the rate high energy modes are eliminated
as far as all modes are eventually eliminated.
Moreover, at each step of mode elimination, 
one could have chosen $\alpha$ differently.
Therefore, $\alpha$ can be regarded as a function of $z$.
If one interprets $z$  as `time', 
it is natural to identify $\alpha(z)$ 
as the `lapse function',
that is, $\alpha(z) = \sqrt{ g_{zz}(z) }$,
where $g_{zz}(z)$ is the metric.
Then one can view Eqs. (\ref{Zgravity}) and (\ref{Sgravity}) as
an one-dimensional gravitational theory with matter fields $J_n$.
This becomes more clear 
if we write the Lagrangian as
\bqa
L = P_n \partial_z J_n - \alpha H,
\eqa
where $H$ is the Hamiltonian
(the reason why $H$ is not Hermitian in this case 
is that we started from the Euclidean field theory).
However, there is one important difference from 
the usual gravitational theory.
In the Hamiltonian formalism of gravity\cite{ADM},
the lapse function is a Lagrangian multiplier
which imposes the constraint $H=0$.
However, in Eq. (\ref{Zgravity}), $\alpha$ is not integrated over
and the Hamiltonian constraint is not imposed.
This is due to the presence of the boundary at $z=0$
which explicitly breaks the reparametrization symmetry.
In particular, the `proper time' 
from $z=0$  to $z=\infty$ given by
\bq 
l = \int_0^\infty \alpha(z) dz
\eq
is a quantity of physical significance
which measures the total warping factor.
To reproduce the original partition function in Eq. (\ref{ZJ})
from Eq. (\ref{Zgravity}),
one has to make sure that $l=\infty$ to include
all modes in the infrared limit.
Therefore, $l$ should be fixed to be infinite.
As a result, one should not integrate over all possible
$\alpha(z)$ some of which give different $l$.
This is the physical reason 
why the Hamiltonian constraint is not imposed
in the present theory.
This theory is a gravitational theory
with the fixed size along the $z$ direction.
Nonetheless, the partition function does not depend on 
a specific choice of $\alpha(z)$ as far as $l$ is fixed.
If one wants to make this gauge symmetry more explicit,
one could integrate in the gauge degree of freedom
by summing over different $\alpha(z)$ with fixed $l$.
For example, we can integrate over $\kappa_{z_1,z_2}$ 
parametrizing $l$-preserving fluctuations of $\alpha$ as
\bqa
\alpha(z) = \alpha_0(z) + \kappa_{z_1,z_2}( \delta(z-z_1) - \delta(z-z_2) ),
\eqa
where $\alpha_0(z)$ is a `gauge fixed' lapse function.
Integration over $\kappa_{z_1,z_2}$ generates the constraint
\bqa
H(z_1) = H(z_2).
\eqa
However, this is a trivial constraint
which is already implemented in Eq. (\ref{Zgravity}) with fixed $\alpha$,
because it is simply the conservation of `energy'.

Although there are many fields in the bulk, i.e. $J_n, P_n$ for each $n$, 
there is only one propagating mode,
and the remaining fields are non-dynamical in the sense they 
strictly obey constraints imposed by their conjugate fields.
This is not surprising because we started with one dynamical field $\Phi$.
There is a freedom in choosing one independent field.
In this case, it is convenient to choose $J_3$ 
as an independent field 
because the conjugate field $P_3$ is multiplied by $J_4$ in Eq. (\ref{Sgravity}),
where $J_4$ is non-dynamical due to Eq. (\ref{J4}).
Integrating over $P_3$, one obtains
\bqa
S & = & \int_0^\infty dz \Bigl[
\frac{M^2}{32 \alpha J_4^2} ( \partial_z J_3 + 3 \alpha J_3 )^2
- \frac{i}{4 J_4} ( \partial_z J_3 + 3 \alpha J_3 ) ( 3 J_3 P_2 + 2 J_2 P_1 + i J_1 ) \nn
&&  ~~~~~~~~~~~ + i (\partial_z J_1 + \alpha J_1 ) P_1
 + i (\partial_z J_2 + 2 \alpha J_2 ) P_2
\Bigr].
\eqa
$P_1$ and $P_2$ are Lagrangian multipliers which impose the constraints,
\bqa
(\partial_z J_1 + \alpha J_1 ) & = & 
\frac{J_2}{2 J_4} ( \partial_z J_3 + 3 \alpha J_3 ), \nn
(\partial_z J_2 +  2 \alpha J_2 ) & = & 
\frac{3 J_3}{4 J_4} ( \partial_z J_3 + 3 \alpha J_3 ).
\eqa
Remarkably, these constraints have a local solution,
that is, the fields $J_1$ and $J_2$ at a scale $z$ 
depend only on the independent field $J_3$ at the same scale,
\bqa
J_1 & = & \frac{J_3^3}{16 J_4^2} + 
\JJ_2 e^{-2 \alpha z} \frac{J_3}{2 J_4}, \nn 
J_2 & = & \frac{3 J_3^2}{8 J_4} + \JJ_2 e^{-2 \alpha z}.
\eqa
This locality is guaranteed because
all source fields at a given scale are
tied with one fluctuating field  $\td \phi$ at the same scale.
Here we considered the case with $Z_2$ symmetry where 
$\JJ_n=0$ for odd $n$.
From this one can write down the local action 
for the independent field $J_3$ in the bulk,
\bqa
S & = & \int_0^\infty dz \left[
\frac{M^2}{32 \alpha J_4^2} ( \partial_z J_3 + 3 \alpha J_3 )^2 
 + \partial_z \left\{
\frac{J_3^4}{256 J_4^3} 
+ \frac{\JJ_2 e^{-2 \alpha z}}{16 J_4^2}
 J_3^2
\right\}
\right].
\label{ZD}
\eqa
The first term is a bulk term
and the second term is a boundary term.
Since $J_3(z=0)=0$,
the boundary terms contribute only at the infrared limit $z=\infty$.
The theory for $J_3$ is free in the bulk,
but the boundary term contains non-trivial interactions.
Presumably, this theory is not easier to solve 
than the original theory due to the boundary interactions.
However, the construction of the dual theory for the toy model
illustrates how one can construct
dual theories for more general field theories.
Now, rather than trying to analyze the theory (\ref{ZD}),
we will move on to apply the prescription
to more non-trivial field theory : 
$D$-dimensional $O(N)$ vector model.

\section{$D$-dimensional O(N) vector theory}

\subsection{Construction of dual theory}

We consider a $D$-dimensional vector field theory,
\bqa
Z[\JJ] & = & \int D \Phi_a ~e^{- ( S_M [\Phi] + S_\JJ[\Phi])},
\label{ONZ}
\eqa
where
\bqa
S_{M}[\Phi] &=& \int d {\bf x} d {\bf y} ~\Phi_a({\bf x}) G_M^{-1}({\bf x}-{\bf y}) \Phi_a({\bf y}), \nn
S_\JJ[\Phi] &=& \int d {\bf x} ~ \Bigl[
\JJ_a \Phi_a + \JJ_{ab} \Phi_a \Phi_b + \JJ_{abc} \Phi_a \Phi_b \Phi_c + \JJ_{abcd} \Phi_a \Phi_b \Phi_c \Phi_d \nn
&& + \JJ_{ab}^{ij}  \partial_i \Phi_a \partial_j \Phi_b
+ \JJ_{abc}^{ij} \Phi_a \partial_i \Phi_b \partial_j \Phi_c
+ \JJ_{abcd}^{ij} \Phi_a \Phi_b \partial_i \Phi_c \partial_j \Phi_d
\Bigr].
\label{ONS}
\eqa
Here $\int d {\bf x}$ 
and $\int d {\bf y}$ 
are integrations 
on a $D$-dimensional manifold ${\cal M}^D$.
Here we use ${\cal M}^D = \mathbb{R}^D$ for simplicity.
$\Phi_a$ is $O(N)$ vector field.
$G_M^{-1}({\bf x})$ is the regularized kinetic energy with
\bqa
G_M^{-1}({\bf x}) & = & \int d {\bf p} ~ p^2 K^{-1}\left( p/M \right) e^{i p x},
\label{green}
\eqa
where $p x \equiv p_i x_i$.
$K^{-1}(s)$ is an analytic function of $s^2$,
which remains to be order of $1$ for $s<1$ 
and grows smoothly for $s>1$, for example,
\bqa
K^{-1}(s) & = & e^{s^2}.
\label{K}
\eqa
The mass scale $M$ is a UV cut-off
above which fluctuations of $\Phi_a$ are suppressed.
$S_\JJ$ is a deformation of the free theory.
We consider sources $\JJ_{ab...}$ 
which are fully symmetric in the flavor indices $a,b,...$.
In general, the sources may depend on ${\bf x}$.
Although we can add more general deformations,  
we will proceed with this quartic action (\ref{ONS})
which is sufficient to illustrate 
general features of the holographic description.

To integrate out high energy modes,
we add an auxiliary vector field $\td \Phi_a$,
\bqa
Z[\JJ] & = & [\det \td G_D]^{-N/2} \int D \Phi D \tilde \Phi ~e^{-(S_M[\Phi] + S_\JJ[\Phi] + \tilde S[\td \Phi])},
\eqa
where
\bqa
\td S[ \td \Phi ] & = & \int d {\bf x} d {\bf y} ~ \td \Phi_a({\bf x}) \td G^{-1}_D({\bf x}-{\bf y}) \td \Phi_a({\bf y}).
\eqa
The form of the propagator $\td G_D$ for the auxiliary field
does not affect the final answer.
Then, we find a new basis $\phi$ and $\tilde \phi$, 
\bqa
\Phi_a({\bf x}) &=& \phi_a({\bf x}) + \tilde \phi_a({\bf x}), \nn
\tilde \Phi_a({\bf x}) &=& \int d {\bf y} ~ \left( A({\bf x},{\bf y}) \phi_a({\bf y}) + B({\bf x},{\bf y}) \tilde \phi_a({\bf y}) \right),
\eqa
where $A$ and $B$ are chosen to satisfy
\bqa
G_M^{-1} + A^T \td G_D^{-1} A & = & G^{-1}_{M^{'}}, \nn
G_M^{-1} + B^T \td G_D^{-1} B & = & \td G^{'-1}, \nn
G_M^{-1} + A^T \td G_D^{-1} B & = & 0
\label{eq:dia}
\eqa
so that the low energy field $\phi$ has a slightly
smaller UV cut-off $M^{'} = M e^{-\alpha dz}$ 
and the high energy field $\td \phi$ has a propagator
$\td G^{'} =  - (G_{M^{'}} - G_M)$. 
Then the partition function can be written as
\bqa
Z[\JJ] 
& = & [\det \td G^{'-1} \det G^{-1}_{M^{'}} \det G_{M}]^{N/2}
 \int D \phi D \tilde \phi  ~e^{-(S_\JJ[\phi+\td \phi] + S_{M^{'}}[\phi] + \tilde S^{'}[\td \phi])},
\eqa
where
\bqa
\td S^{'} & = & \int d {\bf x} d {\bf y} ~ \td \phi_a({\bf x}) \td G^{'-1}({\bf x}-{\bf y}) \td \phi_a({\bf y}).
\eqa
A rescaling of ${\bf x}$ and the fields,
\bqa
{\bf x} &\rightarrow& e^{\alpha dz} {\bf x}, \nn
\phi_a & \rightarrow & e^{ (2-d) \alpha dz /2  } \phi_a,\nn
\td \phi_a & \rightarrow & e^{ (2-d) \alpha dz /2  } \td \phi_a
\label{scale}
\eqa
brings the kinetic energy for the low energy field
to the original form as 
\bqa
Z[\JJ] 
& = & [\det \td G^{-1} ]^{N/2}
 \int D \phi D \tilde \phi  ~e^{-(S_j[\phi+\td \phi] + S_{M}[\phi] + \tilde S[\td \phi])},
\label{Z10}
\eqa
where 
\bqa
S_j[\phi] &=& \int d {\bf x} ~ \Bigl[
j_a \phi_a + j_{ab} \phi_a \phi_b + j_{abc} \phi_a \phi_b \phi_c + j_{abcd} \phi_a \phi_b \phi_c \phi_d \nn
&& + j_{ab}^{ij}  \partial_i \phi_a \partial_j \phi_b
+ j_{abc}^{ij} \phi_a \partial_i \phi_b \partial_j \phi_c
+ j_{abcd}^{ij} \phi_a \phi_b \partial_i \phi_c \partial_j \phi_d
\Bigr]
\eqa
with 
\bqa
j_{a}({\bf x}) &=&  e^{ \frac{2+D}{2} \alpha dz }  ~\JJ_{a}(e^{\alpha dz} {\bf x}), \nn
j_{ab}({\bf x}) &=&  e^{ 2 \alpha dz } ~ \JJ_{ab}(e^{\alpha dz} {\bf x}), \nn
j_{abc}({\bf x}) &=&  e^{ \frac{6-D}{2} \alpha dz } ~ \JJ_{abc}(e^{\alpha dz} {\bf x}), \nn
j_{abcd}({\bf x}) &=&  e^{ (4-D) \alpha dz } ~ \JJ_{abcd}(e^{\alpha dz} {\bf x}), \nn
j^{ij}_{ab}({\bf x}) &=& \JJ^{ij}_{ab}(e^{\alpha dz} {\bf x}), \nn
j^{ij}_{abc}({\bf x}) &=&  e^{ \frac{2-d}{2} \alpha dz } ~ \JJ^{ij}_{abc}(e^{\alpha dz} {\bf x}), \nn
j^{ij}_{abcd}({\bf x}) &=&  e^{ (2-d) \alpha dz } ~ \JJ^{ij}_{abcd}(e^{\alpha dz} {\bf x})
\eqa
and 
\bqa
\td G^{-1}({\bf x}-{\bf y}) = e^{ (2+d) \alpha dz} \td G^{'-1}( e^{\alpha dz}({\bf x}-{\bf y}) ). 
\eqa

The new action can be expanded in power of the low energy field,
\bqa
S_{j}[\phi + \td \phi]  
 & = & 
S_j[ \td \phi] \nn
& & + \int d {\bf x} \Bigl\{  \Bigl[
 j_a  + 2 j_{ab} \td \phi_b + 3 j_{abc} \td \phi_b \td \phi_c + 4 j_{abcd} \td \phi_b \td \phi_c \td \phi_d  \nn
&& ~~ - 2 \partial_i ( j_{ab}^{ij} \partial_j \td \phi_b ) 
+ j_{abc}^{ij} \partial_i \td \phi_b \partial_j \td \phi_c  
- 2 \partial_i( j_{abc}^{ij} \td \phi_b  \partial_j \td \phi_c )  \nn
&& ~~ + 2 j_{abcd}^{ij}  \td \phi_b \partial_i \td \phi_c \partial_j \td \phi_d  
-  2 \partial_i ( j_{abcd}^{ij} \td \phi_b \td \phi_c \partial_j \td \phi_d )
\Bigr] \phi_a \nn
&& + \Bigl[
 j_{ab} + 3 j_{abc} \td \phi_c + 6 j_{abcd} \td \phi_c \td \phi_d  
 - \partial_i ( j_{abc}^{ij}  \partial_j \td \phi_c ) \nn
&&
~~ + j_{abcd}^{ij}  \partial_i \td \phi_c \partial_j \td \phi_d 
- 2 \partial_i (  j_{abcd}^{ij} \td \phi_c  \partial_j \td \phi_d  ) 
\Bigr] \phi_a \phi_b \nn
&& + \Bigl[ j_{ab}^{ij} + j_{abc}^{ij} \td \phi_c + j_{abcd}^{ij} \td \phi_c \td \phi_d \Bigr] \partial_i \phi_a \partial_j \phi_b  \nn
&& + \Bigl[
 j_{abc}  + 4 j_{abcd} \td \phi_d  
 -\frac{2}{3} \partial_i ( j_{abcd}^{ij}  \partial_j \td \phi_d )
\Bigr]\phi_a \phi_b \phi_c \nn
&& + \Bigl[
 j_{abc}^{ij}   + 2 j_{abcd}^{ij} \td \phi_d  
\Bigr]\phi_a \partial_i \phi_b \partial_j \phi_c  \nn
&& + j_{abcd} \phi_a \phi_b \phi_c \phi_d + j_{abcd}^{ij} \phi_a \phi_b \partial_i \phi_c \partial_j \phi_d
\Bigr\}.
\eqa
Here we ignored boundary terms in the $D$-dimensional space,
assuming that the boundary is at infinity 
where couplings vanish.
We decouple the low energy field
and the high energy field 
by introducing Hubbard-Stratonovich fields,
\bqa
Z[\JJ] 
& = & [\det \td G^{-1} ]^{N/2}
 \int D \phi D \tilde \phi DJ DP ~
e^{-(S_j^{'} + S_{M}[\phi] + \tilde S[\td \phi] )},
\eqa
where
\bqa
S_j^{'} & = & S_j[\td \phi] \nn
& =  \int d {\bf x} & 
\Bigl\{  
iP_a J_a - i P_a 
\Bigl[ j_a  + 2 j_{ab} \td \phi_b + 3 j_{abc} \td \phi_b \td \phi_c + 4 j_{abcd} \td \phi_b \td \phi_c \td \phi_d  \nn
&& ~~ - 2 \partial_i ( j_{ab}^{ij} \partial_j \td \phi_b ) 
+ j_{abc}^{ij} \partial_i \td \phi_b \partial_j \td \phi_c  
- 2 \partial_i( j_{abc}^{ij} \td \phi_b  \partial_j \td \phi_c )  \nn
&& ~~ + 2 j_{abcd}^{ij}  \td \phi_b \partial_i \td \phi_c \partial_j \td \phi_d  
-  2 \partial_i ( j_{abcd}^{ij} \td \phi_b \td \phi_c \partial_j \td \phi_d )
\Bigr] + J_a \phi_a \nn
&& + iP_{ab} J_{ab} - i P_{ab} \Bigl[
 j_{ab} + 3 j_{abc} \td \phi_c + 6 j_{abcd} \td \phi_c \td \phi_d  
 - \partial_i ( j_{abc}^{ij}  \partial_j \td \phi_c ) \nn
&&
~~ + j_{abcd}^{ij}  \partial_i \td \phi_c \partial_j \td \phi_d 
- 2 \partial_i (  j_{abcd}^{ij} \td \phi_c  \partial_j \td \phi_d  ) 
\Bigr] + J_{ab} \phi_a \phi_b \nn
&& + iP_{ab,ij} J_{ab}^{ij} - iP_{ab,ij} \Bigl[ j_{ab}^{ij} + j_{abc}^{ij} \td \phi_c + j_{abcd}^{ij} \td \phi_c \td \phi_d \Bigr] 
+ J_{ab}^{ij} \partial_i \phi_a \partial_j \phi_b  \nn
&& + iP_{abc} J_{abc} - iP_{abc} \Bigl[
 j_{abc}  + 4 j_{abcd} \td \phi_d  
 -\frac{2}{3} \partial_i ( j_{abcd}^{ij}  \partial_j \td \phi_d )
\Bigr] + J_{abc} \phi_a \phi_b \phi_c \nn
&& + iP_{abc,ij} J_{abc}^{ij} - iP_{abc,ij} \Bigl[
 j_{abc}^{ij}   + 2 j_{abcd}^{ij} \td \phi_d  
\Bigr] + J_{abc}^{ij} \phi_a \partial_i \phi_b \partial_j \phi_c  \nn
&& + iP_{abcd} J_{abcd} - iP_{abcd} j_{abcd} + J_{abcd} \phi_a \phi_b \phi_c \phi_d \nn
&& + iP_{abcd,ij} J_{abcd}^{ij} - iP_{abcd,ij} j_{abcd}^{ij} + J_{abcd}^{ij} \phi_a \phi_b \partial_i \phi_c \partial_j \phi_d.
\eqa

Now we integrate out $\td \phi$, 
following the similar procedure 
as explained in the Appendix A to obtain 
\bqa
Z[\JJ] & = &    
\int d \phi D J D P  ~ e^{-(S_{J}[\phi] + S_M[\phi] + S^{(1)}[J,P])},
\label{Zd}
\eqa
where
\bqa
S^{(1)}[J,P] & =  \int d {\bf x} & 
\Bigl\{  
i P_a ( \partial J_a - \frac{2+D}{2} \alpha \JJ_a ) 
+i P_{ab} ( \partial J_{ab} - 2 \alpha \JJ_{ab} ) 
+i P_{ab,ij} ( \partial J_{ab}^{ij} )  \nn
&& + i P_{abc} ( \partial J_{abc} - \frac{6-D}{2} \alpha \JJ_{abc} ) 
+i P_{abc,ij} ( \partial J_{abc}^{ij} - \frac{2-d}{2} \alpha \JJ_{abc}^{ij} )  \nn
&& 
+i P_{abcd} ( \partial J_{abcd} - (4-D) \alpha \JJ_{abcd} ) 
+i P_{abcd,ij} ( \partial J_{abcd}^{ij} - (2-d) \alpha \JJ_{abcd}^{ij} )  
\Bigr\} dz \nn
& + \frac{1}{4} \int d {\bf x} d {\bf y} & 
\Bigl\{  
s_a({\bf x}) \td G({\bf x}-{\bf y}) s_a({\bf y})
\Bigr\}.
\label{Sd}
\eqa
Here,
$ \partial = \frac{\partial}{\partial z} - 
\alpha \sum_{i=1}^D  x_i \frac{\partial}{\partial x_i}$
and
\bqa
s_a & = &
\Bigl[
 i \td \JJ_a 
+ 2  P_b \td \JJ_{ab} 
- 2  \partial_j ( \td \JJ_{ab}^{ij} \partial_i P_b )
+ 3  P_{bc} \td \JJ_{abc} 
-  \partial_j ( \td \JJ_{abc}^{ij} \partial_i P_{bc} ) \nn
&& ~~ 
+  P_{bc,ij} \td \JJ_{abc}^{ij}
+ 4  P_{bcd} \td \JJ_{abcd} 
- \frac{2}{3}  \partial_j ( \td \JJ_{abcd}^{ij} \partial_i P_{bcd} )
+ 2  \td \JJ_{abcd}^{ij} P_{bcd,ij}
\Bigr]
\eqa
with $\td \JJ_{a..} = ( \JJ_{a..} + J_{a..})/2$ and
$\td \JJ_{a..}^{ij} = ( \JJ_{a..}^{ij} + J_{a..}^{ij} )/2$.
If one integrate out $J$'s and $P$'s in Eq. (\ref{Zd}),
one reproduces the action obtained by 
integrating out $\td \phi$ in Eq. (\ref{Z10}) 
to the order of $dz$.
If we keep applying the same procedure 
to the action for the low energy field
as we did in the previous section,
the partition function can be written as
\bqa
Z[\JJ] & = &    
\int D J D P ~ e^{-S[J,P]},
\label{Zgravity5}
\eqa
where the $(D+1)$-dimensional action is given by
\bqa
S[J,P] & =  \int d {\bf x} dz & 
\Bigl\{  
i P_a ( \partial J_a - \frac{2+D}{2} \alpha J_a )
+i P_{ab} ( \partial J_{ab} - 2 \alpha J_{ab} )
+i P_{ab,ij} ( \partial J_{ab}^{ij} ) \nn
&& + i P_{abc} ( \partial J_{abc} - \frac{6-D}{2} \alpha J_{abc} )
+i P_{abc,ij} ( \partial J_{abc}^{ij} - \frac{2-d}{2} \alpha J_{abc}^{ij} ) \nn
&& 
+i P_{abcd} ( \partial J_{abcd} - (4-D) \alpha J_{abcd} )
+i P_{abcd,ij} ( \partial J_{abcd}^{ij} - (2-d) \alpha J_{abcd}^{ij} )
\Bigr\} \nn
& + \frac{1}{4} \int d {\bf x} d {\bf y} dz & 
\Bigl\{  
\alpha s_a({\bf x}) G^{'}({\bf x}-{\bf y}) s_a({\bf y})
\Bigr\},
\label{Sgravity5}
\eqa
with
\bqa
s_a & = &
\Bigl[
 i J_a 
+ 2  P_b J_{ab} 
- 2  \partial_j ( J_{ab}^{ij} \partial_i P_b )
+ 3  P_{bc} J_{abc} 
-  \partial_j ( J_{abc}^{ij} \partial_i P_{bc} ) \nn
&& ~~ 
+  P_{bc,ij} J_{abc}^{ij}
+ 4  P_{bcd} J_{abcd} 
- \frac{2}{3}  \partial_j ( J_{abcd}^{ij} \partial_i P_{bcd} )
+ 2  J_{abcd}^{ij} P_{bcd,ij}
\Bigr],
\eqa
and 
$G^{'}({\bf x}) \equiv  M \partial_M G_M({\bf x})$.
Here the partition function is given by the 
functional integrals of the source fields $J$ and their
conjugate fields $P$ in the $(D+1)$-dimensional space 
${\cal M}^D \times [0,\infty)$.
If the $D$-dimensional manifold ${\cal M}^D$ has a finite volume $V$,
the volume at scale $z$ is given by $V e^{-\alpha D z}$
because of the rescaling of space in Eq. (\ref{scale}).
Since $G_M({\bf p})^{-1}$ is smooth in momentum space, 
the last term in Eq. (\ref{Sgravity5}) can be expressed in gradient expansion.
Higher derivative terms are suppressed by $\left( \frac{ \partial_i}{M} \right)^n$.
Therefore the full theory is local in $(D+1)$-dimensional space.
The locality along the emergent $z$ direction is due to the
fact that physics at a given energy scale $E$ depends 
on higher energy physics only through physics
at an infinitesimally higher energy scale $E e^{\alpha dz}$.

As is the case 
in the $0$-dimensional theory 
discussed in the previous section,
$\alpha$ can be regarded as the lapse function
in a gravitational theory in $(D+1)$ dimensions, 
but the Hamiltonian constraint is not imposed 
because of the boundary at $z=0$.
Moreover, there is no fluctuating shift function 
which imposes the momentum constraint $P_i=0$
in the usual gravitation theory.
This is, again, due to fact that 
the boundary explicitly breaks the diffeomorphism symmetry
and only the momentum conservation (not $P_i=0$) 
is implemented in the theory.

One key difference from the $0$-dimensional theory is that
there exist bulk fields with non-trivial spins.
In Eq. (\ref{Sgravity5}), there are spin two fields
which are coupled to the energy momentum tensor at the boundary.
In the presence of more general deformations in the boundary theory,
one needs to introduce fields with higher spins\cite{VASILIEV,PETKOU}.


\subsection{Large $N$ limit}

In this section, we will 
consider a large $N$ limit where
the dual theory becomes classical 
for $O(N)$ singlet fields.
To see this, one decomposes tensors 
with rank two or greater than two
into traceless tensors with the same rank
and tensors with lower ranks as
\bqa
J_{ab}  & = & J_2  \delta_{ab} + \frac{1}{N} \bar J_{ab} , \nn
J_{abc}  & = & \frac{1}{N} \left[ J_{3a}  \delta_{bc} + J_{3b}  \delta_{ac} + J_{3c}  \delta_{ab} \right] + \frac{1}{N^2} \bar J_{abc} , \nn
J_{abcd}  & = & \frac{J_4 }{N} ( \delta_{ab} \delta_{cd} + \delta_{ac} \delta_{bd} + \delta_{ad} \delta_{bc}), \nn
P_{ab}  & = & P_2  \delta_{ab} + \bar P_{ab} , \nn
P_{abc}  & = & P_{3a}  \delta_{bc} + P_{3b}  \delta_{ac} + P_{3c}  \delta_{ab} + \bar P_{abc} , \nn
P_{abcd}  & = & P_4  ( \delta_{ab} \delta_{cd} + \delta_{ac} \delta_{bd} + \delta_{ad} \delta_{bc}), \nn  
J_{ab}^{ij}  & = & J_2^{ij}  \delta_{ab} + \frac{1}{N} \bar J_{ab}^{ij} , \nn
J_{abc}^{ij}  & = & \frac{1}{N} \left[ J_{3a}^{ij}  \delta_{bc} + J_{3b}^{ij}  \delta_{ac} + J_{3c}^{ij}  \delta_{ab} \right]  + \frac{1}{N^2} \bar J_{abc}^{ij} , \nn
J_{abcd}^{ij}  & = & \frac{J_4^{ij} }{N} ( \delta_{ab} \delta_{cd} + \delta_{ac} \delta_{bd} + \delta_{ad} \delta_{bc}), \nn
P_{ab,ij}  & = & P_{2,ij}  \delta_{ab} + \bar P_{ab,ij} , \nn
P_{abc,ij}  & = & P_{3a,ij}  \delta_{bc} + P_{3b,ij}  \delta_{ac} + P_{3c,ij}  \delta_{ab} + \bar P_{abc,ij} , \nn
P_{abcd,ij}  & = & P_{4,ij}  ( \delta_{ab} \delta_{cd} + \delta_{ac} \delta_{bd} + \delta_{ad} \delta_{bc}),
\eqa
where the fields with the bar are traceless,
$\bar J_{aa} = \bar J_{aab} = \bar J_{aa}^{ij} = \bar J_{aab}^{ij} = 0$
and 
$\bar P_{aa} = \bar P_{aab} = \bar P_{aa,ij} = \bar P_{aab,ij} = 0$.
For the quartic couplings which are the highest order couplings,
only the $O(N)$ invariant parts are kept,
assuming that the bare quartic couplings are $O(N)$ invariant.
Note that the structures of the highest order coupling fields
are not modified at all $z$, 
which is not true for other coupling fields :
for lower order couplings, non-singlet 
contributions are generated at low energy scales
even though the bare couplings are $O(N)$ invariant.
Now, we consider the large $N$ limit with fixed
$J_a$, $J_2$, $\bar J_{ab}$, $J_{3a}$, $\bar J_{abc}$, $J_4$,
$J_2^{ij}$, $\bar J_{ab}^{ij}$, 
$J_{3a}^{ij}$, $\bar J_{abc}^{ij}$, $J_4$, $J_4^{ij}$.
In this limit, the leading order action becomes
\bqa
S[J,P] & =  \int d {\bf x} dz & 
\Bigl\{  
i P_a ( \partial J_a - \frac{2+D}{2} \alpha J_a )
+i N P_{2} ( \partial J_{2} - 2 \alpha J_{2} )
+i \frac{1}{N} \bar P_{ab} ( \partial \bar J_{ab} - 2 \alpha \bar J_{ab} ) \nn
&&
+i N P_{2,ij} ( \partial J_{2}^{ij} ) 
+i \frac{1}{N} \bar P_{ab,ij} ( \partial \bar J_{ab}^{ij} ) \nn
&& + i 3P_{3a} ( \partial J_{3a} - \frac{6-D}{2} \alpha J_{3a} ) 
+ i \frac{1}{N^2} \bar P_{abc} ( \partial \bar J_{abc} - \frac{6-D}{2} \alpha \bar J_{abc} ) \nn
&& 
+ i 3 P_{3a,ij} ( \partial J_{3a}^{ij} - \frac{2-d}{2} \alpha J_{3a}^{ij} ) 
+ i \frac{1}{N^2} \bar P_{abc,ij} ( \partial \bar J_{abc}^{ij} - \frac{2-d}{2} \alpha \bar J_{abc}^{ij} ) \nn
&& 
+i 3 N P_{4} ( \partial J_{4} - (4-D) \alpha J_{4} )
+i 3 N P_{4,ij} ( \partial J_{4}^{ij} - (2-d) \alpha J_{4}^{ij} )
\Bigr\} \nn
& + \frac{1}{4} \int d {\bf x} d {\bf y} dz & 
\Bigl\{  
\alpha  s_a({\bf x},z) G^{'}({\bf x}-{\bf y}) s_a({\bf y},z)
\Bigr\},
\eqa
where
\bqa
s_a & = &
\Bigl[
 i J_a 
+ 2  P_a J_2 
+ \frac{2}{N}  P_b \bar J_{ab} 
- 2  \partial_j ( J_{2}^{ij} \partial_i P_a )
- \frac{2}{N}  \partial_j ( \bar J_{ab}^{ij} \partial_i P_b ) \nn
&& + 3  P_{2} J_{3a} 
+ \frac{6}{N} \bar P_{ab} J_{3b}
+ \frac{3}{N^2}  \bar P_{bc} \bar J_{abc} \nn
&&
-   \partial_j ( J_{3a}^{ij} \partial_i P_{2} ) 
- \frac{2}{N}  \partial_j ( J_{3b}^{ij} \partial_i \bar P_{ba} ) 
- \frac{1}{N^2} \partial_j ( \bar J_{abc}^{ij} \partial_i \bar P_{bc} ) \nn
&& ~~ 
+   P_{2,ij} J_{3a}^{ij}
+  \frac{2}{N} \bar P_{ab,ij} J_{3b}^{ij}
+  \frac{1}{N^2} \bar P_{bc,ij} \bar J_{abc}^{ij} \nn
&& + 12  P_{3 a} J_{4} 
- 2   \partial_j ( J_{4}^{ij} \partial_i P_{3a} )
+ 6   J_{4}^{ij} P_{3a,ij}
\Bigr].
\eqa

In the large $N$ limit, 
the action is manifestly proportional to $N$.
Therefore, 
one can ignore quantum fluctuations 
for singlet fields such as $J_2$ and $J_2^{ij}$,
and it is enough to consider saddle point solutions
for singlet fields to compute correlation functions 
of singlet operators.
It would be natural 
to integrate out all non-singlet fields
and obtain an effective theory for single fields alone.
However, it turns out that 
the effective action for single fields
become non-local in this $O(N)$ vector model.
This is because there are light non-singlet fields
in the bulk and integrating over those soft modes
generates non-local correlations for singlet fields.
This means that we should keep light non-singlet fields
as `low energy degrees of freedom' in the bulk description 
if we want to use a local description.

\subsection{Phase Transition and Critical Behaviors }

In this section, we will describe 
the phase transition
and the critical properties of the model in $D>2$
using the holographic theory.
To maintain the locality of the description, 
we will keep light non-singlet fields in the theory
by choosing $J_{3a}$'s as independent fields.
We will focus on 
the simple case 
where there is no deformation on the energy-momentum tensor,
and all sources respect the O(N) symmetry,
\bqa
J_a({\bf x},0)  &=&  0, \nn
J_{ab}({\bf x},0) &=& \JJ_{2}({\bf x}) \delta_{ab}, \nn
J_{abc}({\bf x},0) &=& 0, \nn
J_{abcd}({\bf x},0) &=& \frac{\JJ_{4}({\bf x})}{N} \left( 
\delta_{ab}\delta_{cd}
+\delta_{ac}\delta_{bd}
+\delta_{ad}\delta_{bc} 
\right)
\label{bc}
\eqa
with 
$J_2^{ij} = \bar J_{ab}^{ij} =J_{3a}^{ij} = \bar J_{abc}^{ij} = J_{4}^{ij} = 0$.
From now on, it will be assumed that 
$\JJ_4({\bf x})=\JJ_4 > 0$ is $x$-independent,
but $\JJ_2({\bf x})$, 
which may have either sign, 
is $x$-dependent in general.
We first integrate over $P_{3a}$ 
to obtain
\bqa
S[J,P] & = & \int d {\bf x} dz ~~ 
\Bigl\{  
i P_a ( \partial J_a - \frac{2+D}{2} \alpha J_a )
+i N P_{2} ( \partial J_{2} - 2 \alpha J_{2} )
+i \frac{1}{N} \bar P_{ab} ( \partial \bar J_{ab} - 2 \alpha \bar J_{ab} ) \nn
&& 
 ~~~~~~~~~~
 -  i \frac{s_a^{'}}{4 J_4} ( \partial J_{3a} - \frac{6-D}{2} \alpha J_{3a} ) 
+ i \frac{1}{N^2} \bar P_{abc} ( \partial \bar J_{abc} - \frac{6-D}{2} \alpha \bar J_{abc} ) \nn
&& 
~~~~~~~~~~
+i 3 N P_{4} ( \partial J_{4} - (4-D) \alpha J_{4} )
\Bigr\} \nn
&&
 +  \frac{1}{16} \int d {\bf x} d {\bf y} dz  ~~
  C({\bf x},z)
 [ \alpha G^{'}({\bf x}-{\bf y}) ]^{-1}
C({\bf y},z)
\Bigr\},
\label{SPT}
\eqa
where $C({\bf x},z) \equiv 
\frac{1}{J_4({\bf x},z)}
\left( \partial J_{3a}({\bf x},z) - \frac{6-D}{2} \alpha J_{3a}({\bf x},z) \right)
$ and
\bqa
s_a^{'} & = &
\Bigl[
 i J_a 
+ 2  P_a J_2 
+ \frac{2}{N}  P_b \bar J_{ab} 
 + 3  P_{2} J_{3a} 
+ \frac{6}{N} \bar P_{ab} J_{3b}
+ \frac{3}{N^2}  \bar P_{bc} \bar J_{abc} 
\Bigr].
\eqa
Now we integrate over 
$P_a, P_2, \bar P_{ab}, \bar P_{abc}, P_4$
to obtain constraints,
\bqa
\left( \partial J_a - \frac{2+D}{2} \alpha J_a \right) 
- \frac{1}{4 J_4} \left( \partial J_{3b} - \frac{6-D}{2} \alpha J_{3b} \right) 
\left( 2 J_2 \delta_{ab} + \frac{2}{N} \bar J_{ab} \right)
& = & 0, \nn
N ( \partial J_2 - 2 \alpha J_2 ) 
- \frac{3}{4 J_4} \left( \partial J_{3a} - \frac{6-D}{2} \alpha J_{3a} \right) J_{3a}
& = & 0, \nn
 \frac{1}{N} ( \partial \bar J_{ab} - 2 \alpha \bar J_{ab} )
- \frac{3}{4 N J_4} \left( \partial J_{3a} - \frac{6-D}{2} \alpha J_{3a} \right) J_{3b} \nn
- \frac{3}{4 N J_4} \left( \partial J_{3b} - \frac{6-D}{2} \alpha J_{3b} \right) J_{3a} 
- \frac{3}{4 N^2 J_4} \left( \partial J_{3c} - \frac{6-D}{2} \alpha J_{3c} \right) \bar J_{abc}
& = & 0, \nn
\partial \bar J_{abc} - \frac{6-D}{2} \alpha \bar J_{abc} 
& = & 0 , \nn
\partial J_{4} - (4-D) \alpha J_{4} 
& = & 0. 
\eqa
For the boundary condition (\ref{bc}),
the solution for the constraints is given by
\bqa
J_a({\bf x},z) & = & 
\frac{3}{16 N J_4({\bf x},z)^2} J_{3a}({\bf x},z) J_{3b}({\bf x},z) J_{3b}({\bf x},z)
+ \frac{J_{3a}({\bf x},z) \JJ_{2}(e^{\alpha z} {\bf x}) e^{2 \alpha z}}{2J_4({\bf x},z)}, \nn
J_2({\bf x},z) & = & 
\frac{3}{8NJ_4({\bf x},z)} J_{3a}({\bf x},z) J_{3a}({\bf x},z)  
+ \JJ_{2}(e^{\alpha z} {\bf x}) e^{2 \alpha z}, \nn
\bar J_{ab}({\bf x},z) & = & \frac{3}{4J_4({\bf x},z)} J_{3a}({\bf x},z) J_{3b}({\bf x},z), \nn
\bar J_{abc}({\bf x},z) & = & 0, \nn
J_4({\bf x},z) & = & \JJ_{4} e^{ (4-D) \alpha z},
\eqa
where all repeated indices are summed as usual.
Using this result,
one can eliminate all fields 
except for $J_{3a}$ to obtain the action,
\bqa
S[J,P] & = & \int d {\bf x} dz  ~~
\partial_z 
\left[
 \frac{ \JJ_2(e^{\alpha z}{\bf x}) e^{2 \alpha z} }{16 J_4^2 } 
J_{3a} J_{3a} 
 +\frac{ 3 \left( J_{3a} J_{3a} \right)^2}{256 N J_4^3}
\right] \nn
&& +  \frac{1}{16} \int d {\bf x} d {\bf y} dz  
  C({\bf x},z)
 [ \alpha G^{'}({\bf x}-{\bf y}) ]^{-1}
C({\bf y},z)
\Bigr\}.
\label{SPT2}
\eqa
In the first line of the above expression,
all fields which do not have an explicit argument
are understood to be at $({\bf x},z)$.
The first term is a boundary action
and the second term is the bulk action.
In the bulk, $J_{3a}$ is massless and its action is purely quadratic.
Although the bulk theory is non-interacting,
the boundary term has non-trivial interactions
and quantum fluctuations for the non-singlet field
are still important.

\begin{figure}
        \includegraphics[height=7cm,width=15cm]{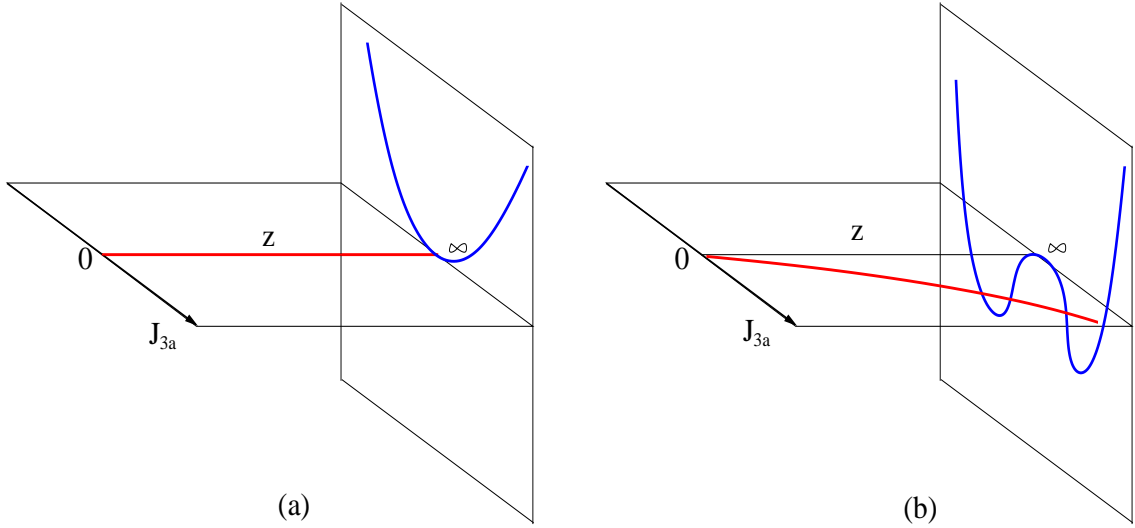} 
\caption{
Saddle point configuration of $J_{3a}(z)$ 
(a) in the disordered phase 
and (b) in the ordered phase.
Although the action is quadratic in $J_{3a}$ in the bulk,
the boundary action drives the phase transition.
When $\JJ_2$ is sufficiently negative,
a Mexican-hat potential at the boundary 
drags $J_{3a}(z)$ away from $J_{3a}(z)=0$ in the bulk.
At the critical point, $J_{3a}$ at the IR boundary $z=\infty$
is more or less free to fluctuate, 
generating algebraic correlations between 
fields inserted at the UV boundary $z=0$.
}
\label{fig:1}
\end{figure}

Due to the boundary term, 
the theory has a non-trivial phase transition.
In the boundary term, the only contribution is from 
the infrared boundary, $z_{I} = \infty$
because there is no symmetry breaking source
in the bare theory, $J_{3a}({\bf x},0)=0$.
To discuss the phase transition,
it is enough to consider $x$ independent singlet source
fields $\JJ_2$ and $\JJ_4$.
If  $\JJ_2 >  \JJ_2^c$, 
where $\JJ_2^c \sim -\JJ_4 M^{(D-2)}$ is a critical value,
the boundary potential has the minimum at $J_{3a}=0$
as is shown in Fig. \ref{fig:1} (a).
Therefore, the minimum energy configuration of $J_{3a}$ 
is the straight line along the $z$-direction.
On the other hand, 
if $\JJ_2 <  \JJ_2^c$
the boundary term has the Mexican-hat potential.
This causes the bulk field $J_{3a}$ to deviate
from the trivial configuration $J_{3a}=0$,
leading to the spontaneous symmetry breaking\cite{KLEBANOV99}
as is shown in Fig. \ref{fig:1} (b).
This describes the second order phase transition
from the disordered phase
to the ordered phase.
We can use $-i P_{3a} \sim \phi^2 \phi_a$ as an order parameter
which roughly measures the slope of $J_{3a}(z)$ along the $z$ direction.
In the phase with broken symmetry,
$J_{3a}$ in the bulk is proportional to $\sqrt{N}$
and fluctuations away from the saddle point is $1/N$ suppressed,
leading to the mean-field like critical exponent for 
the order parameter, $\beta = 1/2$ in the large $N$ limit.

Now let us compute the two-point correlation function of 
the $O(N)$ singlet operator $\phi^2$ at the critical point.
For this we integrate over $J_{3a}$ in the bulk
in the presence of $x$-dependent source 
$\JJ_2({\bf x}) = \JJ^c_{2} + \JJ_2^{'}({\bf x})$
with the boundary condition
$J_{3a}({\bf x},0)  =  0$.
Since the bulk action is quadratic 
one can use the saddle point solution in the bulk.
In terms of the Fourier modes,
\bqa
J_{3a}({\bf x},z) = \int d {\bf p}~~ f_a({\bf p},z) e^{i p x e^{\alpha z} + (6-D)\alpha z/2 },
\eqa
the bulk action becomes
\bqa
S_{bulk} & = & \frac{1}{16 \alpha \JJ_4^2}
\int dz d {\bf p} ~~ e^{-2 \alpha z} [G^{'}]^{-1}( p e^{\alpha z} ) 
[ \partial_z f_a({\bf p},z) ]  
[ \partial_z f_a(-{\bf p},z) ].
\eqa
The equation of motion for $f_{a}({\bf p},z)$ is
\bqa
\partial_z \left( e^{-2 \alpha z} G^{' -1}(p e^{\alpha z}) \partial_z f_a({\bf p},z) \right) = 0.
\eqa
There are two independent solutions : $1$ and $K(p e^{\alpha z} / M)$. 
The boundary condition at $z=0$ fixes 
only one coefficient, 
leading to one-parameter family of solutions
given by
\bqa
f_a({\bf p},z) & = & y_a({\bf p}) \left[ 1 - 
\frac{K\left( p e^{\alpha z} / M \right)}{K\left( p / M \right)}
\right].
\label{ss}
\eqa
Here $K(s)$ is the regulator defined in Eq. (\ref{K}), and
$y_a({\bf p})$ is an arbitrary function.
This arbitrariness is due to the fact that
we have not imposed the boundary condition at the IR limit $z=\infty$.
The condition of finiteness of the action alone does not 
fix the IR boundary condition because both independent solutions
are regular in the IR limit.
Here one should integrate over $y_a$ 
because $y_a$ is also a dynamical field.
In this way, 
the boundary condition in the $IR$ limit
is dynamically determined by the boundary action, 
whereas 
the boundary condition in the $UV$
is fixed by the bare coupling.
Using the saddle point solution, 
one obtains an action for the boundary field $y_a$,
\bqa
S & = & \frac{1}{16 \JJ_4^2} \int d {\bf p} ~~ p^2 K^{-1}(p/M) | y_a({\bf p}) |^2  \nn
&& 
+ \frac{1}{16 \JJ_4^2} \int d {\bf p}_1 d {\bf p}_2 ~~ 
\left[ \JJ_2^c \delta({\bf p}_1+{\bf p}_2) 
+ \JJ_2^{'}(-{\bf p}_1-{\bf p}_2)  \right] 
y_a({\bf p}_1) y_a({\bf p}_2) \nn
&& 
+ \frac{3}{256 N \JJ_4^3} \int d {\bf p}_1 d {\bf p}_2 d {\bf p}_3 d {\bf p}_4 ~~ 
\delta({\bf p}_1+{\bf p}_2+{\bf p}_3+{\bf p}_4) 
y_a({\bf p}_1) y_a({\bf p}_2) y_a({\bf p}_3) y_a({\bf p}_4).
\eqa
Remarkably, this action has the same form as the original action for $\phi_a$, 
although this action is for the source field at the IR boundary.
Despite the fact that 
it is no more convenient to use this dual description 
than using the conventional field theoretic method for the $O(N)$ model,
one can proceed to integrate out the boundary field $y_a$,
by summing over the usual chain diagrams, 
to obtain an effective potential $W[\JJ_2^{'}({\bf q})]$
for the singlet field.
From this, one can readily compute the correlation function.
For example, in $D=3$ one obtains
\bqa
\frac{\partial^2 W}{\partial \JJ_2^{'}({\bf q}) \partial \JJ_2^{'}(-{\bf q})} \sim |{\bf q}| 
+ \mbox{(analytic terms)}
\eqa
This leads to the scaling dimension 
$[\phi^2] = 2$ 
as expected.

The fact that one has to solve essentially the same field theory 
to dynamically impose the IR boundary condition in the dual description 
is somewhat disappointing from the perspective of duality.
However, this may have been expected in the $O(N)$ model.
This is because all $N$ independent fields are massless,
and we kept all of them in the local bulk effective action.
Since no dynamical field has been integrated out,
no information of the original theory has been coarse-grained and
the bulk theory carries the exactly same amount of dynamical information
as the original theory.
An alternative approach would be 
to include more singlet fields such as
$J_{2n}$ associated with the operators $(\phi_a^2)^n$ for $n=1,2,...,N$,
and integrate out all non-singlet fields to obtain
an effective action for $N$ independent singlet fields. 
However, if one keeps only a small number of singlet fields 
and integrate out the remaining fields,
the resulting action will not be local 
because their masses in the bulk are more or less equally spaced\cite{SEAN}.

The holographic description for the $O(N)$ model is not very useful for practical purpose.
After all, we can solve it in a much simpler way.
We believe that the holographic approach 
may become more useful for more strongly interacting theories
where there exist a gap between a small set of operators
with small scaling dimensions
and a large set of operators with large scaling dimensions.
In such models, it is expected that 
quantum fluctuations of heavy propagating modes in the bulk
will be encoded in local effective actions for light modes
which carry dynamical information 
only for operators with small scaling dimensions.
Then, imposing the IR boundary condition for the light modes
may become more tractable than solving the 
original field theory.

\section{Summary and Discussion}

In the present paper, we provided a general prescription
to construct holographic theory for 
general quantum field theory.
Through an explicit construction,
we showed that the holographic description
for the $O(N)$ vector model
correctly reproduce the spontaneous symmetry breaking
phase transition and critical behaviors,
as predicted by standard field theory methods.
In the future, 
it is of great interest to apply this method to
other strongly coupled quantum field theories
where standard field theory techniques fail, 
such as matrix models\cite{tHOOFT,Gross:1990ub,BERENSTEIN}
and non-Fermi liquids in 2+1 dimensions
which are expected to be dual to certain matrix models\cite{SLEE09}.

If general quantum field theories in $D$ dimensions
are holographically dual to certain $(D+1)$-dimensional local theories,
the next question would be 
``What $(D+1)$-dimensional theories
are dual to local $D$-dimensional quantum field theories ?''
If one knows the answer to this question,
one may use the holographic description
to define quantum field theories
and to classify them. 
To answer this question,
it may be helpful to
fully understand general structures
behind the present holographic 
construction.

\section{Acknowledgment}

The author thanks
Allan Adams, 
Guido Festuccia,
Sean Hartnoll, 
Igor Klebanov,
Patrick Lee, 
Hong Liu, 
Joe Polchinski,
T. Senthil, 
and 
Xiao-Gang Wen
for illuminating discussions and comments.
This research was supported by NSERC.
The author also thanks 
KITP for its hospitality and the support
through NSF Grant No. PHY05-51164
during the workshop, 
{\it Quantum criticality and the AdS/CFT correspondence}.

\section{Appendix A}

The action in Eq. (\ref{Si}) can be expanded in power of the
high energy field as
\bqa
S_j^{'} &= &   S_{J}[\phi] + \sum_{n=1}^4 i ( J_n - j_n ) P_n  \nn
&& +  ( j_1 - 2 i P_1 j_2 - 3 i P_2 j_3 - 4 i P_3 j_4 ) \td \phi \nn
&& +  ( j_2 - 3 i P_1 j_3 - 6 i P_2 j_4 ) \td \phi^2 \nn
&& +  ( j_3 - 4 i P_1 j_4 ) \td \phi^3 \nn
&& +  j_4 \td \phi^4.
\label{Si2}
\eqa
Integrating over $\td \phi$ to the order of $1/m^2$, 
one obtains
\bqa
Z[J] & = &  
\int d \phi \Pi_{n=1}^4 (d J_n d P_n ) ~ \frac{m}{\sqrt{m^2 + B}} 
e^{- \left( M^2 \phi^2 + S_{J}[\phi]  + i  \sum_{n=1}^4 ( J_n - j_n ) P_n + \frac{A^2}{4(m^2+B)} \right) },
\label{eq:dz2}
\eqa
where
\bqa
A & = &  ( i j_1 + 2  P_1 j_2 + 3  P_2 j_3 + 4  P_3 j_4 ), \nn
B & = & ( j_2 - 3 i P_1 j_3 - 6 i P_2 j_4 ).
\eqa
The cubic and higher order terms in $\td \phi$
do not contribute to the linear order in $1/m^2 \sim dz$.
If we keep only those terms that are linear in $dz$ in Eq. (\ref{eq:dz2}),
we obtain the action,
\bqa
S = M^2 \phi^2 + S_{J}[\phi]  + i  \sum_{n=1}^4 ( J_n - j_n ) P_n + \frac{A^2}{4m^2} + \frac{B}{2m^2}.
\label{wrong}
\eqa
However, it is not easy to take the continuum limit ($dz \rightarrow 0$)
in this expression for the following reason.
We can regard $j_n$ and $J_n$ 
as being defined at coordinates 
$z$ and $z+dz$ respectively,
where $z$ is the logarithmic energy scale 
in the renormalization group flow.
Then $P_n$ is defined in the interval (or at the middle point of the interval), $[z, z+dz]$.
Usually, $A^2 dz$ can be interpreted as the integration
of $A^2$ between $z$ and $z+dz$ in the continuum limit.
This would be the correct  
if $A$ were a fixed constant of the order of $1$.
In the present case, however, $A$ contains the dynamical field
$P_n$ whose typical amplitude is order of $m \sim \frac{1}{\sqrt{dz}}$.
Therefore, an error of order of $1/m$ in $j_n$ in the integrand
gives a non-trivial contribution to the integration, 
leading to a discrepancy
between the result in Eq. (\ref{wrong}) and the one
obtained in the naive continuum limit.

To fix this problem, we consider the following trick.
First we absorb the 
factor  $\frac{1}{(m^2+B)}$ in front of $A^2$ in the action
into the measure of $P_3$;
we change the variable $P_3$ to $P_3^{'}$ 
in Eq. (\ref{eq:dz2}),
\bqa
P_3 & = & \frac{\sqrt{m^2 + B}}{m} P_3^{'} + 
\frac{ \sqrt{m^2 + B} - m }{ 4 m j_4} (  i j_1 + 2  P_1 j_2 + 3  P_2 j_3 ),
\eqa
which leads to 
\bqa
Z[J] & = &  
\int d \phi \Pi_{n\neq3} (d J_n d P_n ) d J_3 d P_3^{'}  ~  
e^{- \left( M^2 \phi^2 + S_{J}[\phi]  + S^{'}  \right) },
\label{eq:dz3}
\eqa
where
\bqa
S^{'} & = & i \sum_{n \neq 3} ( J_n - j_n ) P_n 
+ \frac{A^{'2}}{4 m^2} \nn
&& 
+ i ( J_3 - j_3 ) 
\left(
\frac{\sqrt{m^2 + B}}{m} P_3^{'} + 
\frac{ \sqrt{m^2 + B} - m }{ 4 m j_4} (  i j_1 + 2  P_1 j_2 + 3  P_2 j_3 )
\right)
\eqa
with $A^{'} = (i j_1 + 2  P_1 j_2 + 3  P_2 j_3 + 4  P_3^{'} j_4)$.
If we take the leading order terms, 
the above expression becomes
\bqa
S^{'} & = & i \sum_{n \neq 3} ( J_n - j_n ) P_n 
+ \frac{A^{'2}}{4 m^2} + i ( J_3 - j_3 ) P_3^{'}.
\eqa
Dropping the prime sign in $P_3$, 
the partition function becomes
\bqa
Z[J] & = &  
\int d \phi \Pi_{n=1}^4 (d J_n d P_n ) ~  
e^{- S^{''} },
\label{eq:dz4}
\eqa
where
\bqa
S^{''} & = &  
M^2 \phi^2 + S_{J}[\phi]  
+  \sum_{n=1}^4 i ( J_n - j_n ) P_n 
+ \frac{1}{4 m^2}   ( i j_1 + 2  P_1 j_2 + 3  P_2 j_3 + 4  P_3 j_4 )^2.
\label{Spp}
\eqa
However, this expression is not completely satisfactory either.
If one integrates out $J_n$ and $P_n$ in this expression,
one obtains an action for $\phi$ which is different from
the result one obtains after integrating out
$\tilde \phi$ directly from Eq. (\ref{expansion})
to the order of $dz$\cite{GUIDO}.
The difference is the contribution from the tadpole diagram.
The tadpole diagram simply shifts the local couplings,
and it turns out that its contribution can be accounted for by replacing 
$j_n$ with $(j_n + J_n)/2$ in the last term of Eq. (\ref{Spp}) as
\bqa
S^{'''} & = &  
M^2 \phi^2 + S_{J}[\phi]  
+  \sum_{n=1}^4 i ( J_n - j_n ) P_n 
+ \frac{1}{4 m^2}   ( i \td j_1 + 2  P_1 \td j_2 + 3  P_2 \td j_3 + 4  P_3 \td j_4 )^2,
\label{Sppp}
\eqa
where $\td j_n = (j_n + J_n)/2$.
Although this is an infinitesimal change,
it still gives a non-trivial contribution 
because $(J_n-j_n) \sim O(1/m)$ and $P_n (J_n-j_n) \sim O(1)$.
It is straightforward to show that with this action
one reproduces the same action as the one
obtained by integrating out
$\tilde \phi$ directly from Eq. (\ref{expansion})
to the order of $dz$.
In Eq. (\ref{Sppp}), the mean value of $j_n$ and $J_n$ 
is multiplied to $P_n$.
Just as the error of the trapezoidal method 
in the usual discrete 
integration is suppressed to $dz^3$,
the difference between Eq. (\ref{Sppp})
and the integration in the continuum limit
becomes sub-leading in $dz$ even though
$P_n \sim 1/\sqrt{dz}$.
Therefore, Eq. (\ref{Sppp}) can be readily
extended to the continuum limit.
If we use $j_n = \JJ_n e^{-n \alpha dz}$ 
and keep the leading order term,
we obtain Eqs. (\ref{eq:dz}) and (\ref{S1}).
It is noted that if one uses Eq. (\ref{wrong})
and take the naive continuum limit, 
some couplings are spuriously shifted
due to the amplified error introduced
in the continuum limit.


\begin{thebibliography}{99}
\bibitem{MALDACENA} J. M. Maldacena, Adv. Theor. Math. Phys. {\bf 2},  231 (1998).
\bibitem{GUBSER} S. S. Gubser, I. R. Klebanov and A. M. Polyakov, Phys. Lett. B {\bf 428}, 105 (1998).
\bibitem{WITTEN} E. Witten, Adv. Theor. Math. Phys. {\bf 2}, 253 (1998).
\bibitem{KLEBANOV} I.R. Klebanov and A.M. Polyakov, Phys. Lett. B {\bf 550}, 213 (2002). 
\bibitem{DAS} S. R. Das and A. Jevicki, Phys. Rev. D {\bf 68}, 044011 (2003).
\bibitem{Gopakumar:2004qb}
  R.~Gopakumar,
  Phys.\ Rev.\  D {\bf 70}, 025009 (2004);
{\it ibid.} {\bf 70}, 025010 (2004).
\bibitem{POLCHINSKI09} I. Heemskerk, J. Penedones, J. Polchinski and J. Sully, 
J. High Energy Phys. {\bf 10}, 079 (2009).
\bibitem{POLCHINSKI84} J. Polchinski, Nucl. Phys. B {\bf 231}, 269 (1984).
\bibitem{POLONYI} For a review, see J. Polonyi, arXiv:hep-th/0110026v2.
\bibitem{VERLINDE} J. de Boer, E. Verlinde and H. Verlinde, 
J. High Energy Phys. {\bf 08}, 003 (2000).
\bibitem{ADM} R. Arnowitt, S. Deser, and C. Misner, Phys. Rev. {\bf 116}, 1322 (1959).
\bibitem{VASILIEV} M. A. Vasiliev, arXiv:hep-th/9910096. 
\bibitem{PETKOU} A. C. Petkou, J. High Energy Phys. {\bf 03}, 049 (2003).
\bibitem{KLEBANOV99} I. R. Klebanov and E. Witten, Nucl. Phys. B {\bf 556}, 89 (1999).
\bibitem{SEAN} I thank Sean Hartnoll for pointing this out to me.
\bibitem{tHOOFT} G. 't Hooft, Nucl. Phys. B {\bf 72}, 461 (1974).
\bibitem{Gross:1990ub}
  D.~J.~Gross and I.~R.~Klebanov,
  Nucl.\ Phys.\  B {\bf 344}, 475 (1990).
\bibitem{BERENSTEIN} D. E. Berenstein, M. Hanada and S. A. Hartnoll, J. High Energy Phys. {\bf 02}, 010 (2009).
\bibitem{SLEE09} S.-S. Lee, Phys. Rev. B {\bf 80}, 165102 (2009).
\bibitem{GUIDO} I thank Guido Festuccia for pointing this out to me.
\end{thebibliography}
\end{document}